\documentclass[preprint]{elsarticle}

\usepackage{setspace}
\usepackage{multirow}\usepackage{pbox}
\usepackage{lineno,hyperref}
\modulolinenumbers[5]
\usepackage[dvipsnames]{xcolor}
\usepackage{changes}

\makeatletter
\@namedef{Changes@AuthorColor}{red}
\colorlet{Changes@Color}{red}
\makeatother

\journal{Journal of Magnetism and Magnetic Materials}

\begin{document}

\begin{frontmatter}

\title{Magnetic characterization of oblique angle deposited Co/CoO on gold nanoparticles}

\author[mymainaddress]{Johanna K. Jochum \corref{mycorrespondingauthor}}
\cortext[mycorrespondingauthor]{Corresponding author \\\textit{present address:} Heinz Maier-Leibnitz Zentrum (MLZ), Technische Universität München, Lichtenbergstr. 1, 85748 Garching, Germany; Bayrisches Geoinstitut, Universität Bayreuth, 95440 Bayreuth, Germany \\ \textit{Email address:} jjochum@frm2.tum.de (J. K. Jochum)}

\author[mysecondaryaddress]{Thomas Saerbeck}
\author[thirdaddress]{Vera Lazenka}
\author[thirdaddress]{Vincent Joly}
\author[fourthaddress]{Lianchen Shan}
\author[fourthaddress]{Hans-Gerd Boyen}
\author[thirdaddress]{Kristiaan Temst}
\author[thirdaddress]{André Vantomme}
\author[mymainaddress]{Margriet J. Van Bael}

\address[mymainaddress]{KU Leuven, Laboratorium voor Vaste-Stoffysica en Magnetisme, Celestijnenlaan 200D, 3001 Leuven, Belgium}
\address[mysecondaryaddress]{Institut Laue-Langevin, 71 Avenue des Martyrs, 38000 Grenoble, France}
\address[thirdaddress]{KU Leuven, Instituut voor Kern- en Stralingsfysica, Celestijnenlaan 200 D, 3001 Leuven, Belgium}
\address[fourthaddress]{Hasselt University, Institute for Materials Research (IMO-IMOMEC),  Wetenschapspark 1, 3590 Diepenbeek,
Belgium}

\begin{abstract}
The influence of a patterned substrate on obliquely deposited, exchange biased Co/CoO films was studied. It was found that substrates decorated with nanoparticle patterns provide the option to manipulate the orientation of the magnetic easy axis in obliquely deposited thin films.
The complementary methods of SQUID magnetometry and polarized neutron reflectometry were used to disentangle the different contributions to the magnetic hysteresis of such complex magnetic systems.
\end{abstract}

\begin{keyword}
exchange bias\sep polarized neutron reflectometry\sep nanoparticles\sep magnetometry\end{keyword}
\end{frontmatter}

%\linenumbers

\section{Introduction}

Tailoring the magnetic anisotropies of a nanoscale system by means of nanostructuring is an interesting way of manipulating its magnetic properties such as coercivity and exchange bias (EB) \cite{Sort2004, Popova2005, Temst2006, Dobrynin2006, Nogues2005, Kovylina2009, Baltz2005}. In a thin-film system, several contributions to the overall anisotropy exist, all with different strength, leading to a complex and tunable phase diagram of the magnetic switching behavior. The reduced dimensionality of a thin film system leads to a shape anisotropy that often causes the magnetic easy axis to lie in the film plane. In-plane unidirectional anisotropy can be induced, for example, by combining ferromagnetic and antiferromagnetic materials, which leads to the well known exchange bias effect \cite{Nogues2005, Nogues1999, Manna2014, MagHet2008} . In addition, uniaxial anisotropy can be induced by oblique angle deposition (OAD), with a strength that depends on film thickness and deposition angle \cite{Lisfi2002, Schlage2016, Quiros2014, Alameda1996}. OAD on nanostructured templates can furthermore induce ordered growth on the nanoscale and lead to nanoparticle formation, whose orientation and size can be expected to show an additional effect on the global anisotropy and other magnetic properties such as exchange bias. 

In this paper, we study the magnetic anisotropy in exchange biased thin films deposited under an oblique angle on homogeneous and Au nanoparticle decorated substrates. The Au nanoparticles (NPs) form an ordered array and promote Co-nanoparticle growth in addition to lateral structuring of the Co film. This leads to a complex magnetic system with potentially competing anisotropies originating from oblique angle deposition, nanostructuring and exchange bias. Using the complementary techniques of SQUID magnetometry and polarized neutron reflectometry (PNR) we attempt to disentangle the different contributions to the magnetic hysteresis of this system. For the sample with NPs we find an in-plane magnetic anisotropy along the deposition direction, for the sample without NPs no preferential magnetization direction could be established.  

The EB in the prototypical Co/CoO system \cite{Meiklejohn1957} manifests itself as a shift in the hysteresis loop along the applied field axis, an increase in coercive field, asymmetric hysteresis loops and a training effect \cite{Meiklejohn1957, Nogues1999, Menendez2013}. The hysteresis loop shift occurs opposite to the direction of the applied field during the field cooling process needed to set the preferential orientation of the antiferromagnet. 
The EB in Co/CoO is among the strongest reported, entirely dominating the magnetocrystalline anisotropy.

For magnetic systems deposited under an oblique angle the orientation of the magnetic easy axis strongly depends on the deposition angle $\phi$ (here we define $\phi$ as the angle between the incoming particle beam and the surface normal see further fig. \ref{fig:Fig1}a)) as well as the layer thickness $t$ \cite{Lisfi2002}.
For large $\phi$ and small thicknesses $t$, the magnetic easy axis tends to lie in the transverse direction. 
This is due to anisotropies in the density that stem from shadowing processes leading to a reduction in compactness of the film in the longitudinal direction compared to the transverse direction \cite{Quiros2014}.
When the thickness is increased for large $\phi$, such as the angle used in this study, the easy axis will first switch from the transverse in-plane direction to the longitudinal in-plane direction. 
Increasing the thickness further, the shape anisotropy originating from the pillar-structure of thick OAD films will start to dominate and slowly lead to a tilt of the magnetic easy axis out of the sample plane and towards the plane of the growth direction of the pillars \cite{Quiros2014, Alameda1986, Schlage2016, Phuoc2013}.

\section{Experimental Methods}

\begin{figure}
\vspace{-1.5cm}
	\centering
		\includegraphics[width=01.0\textwidth]{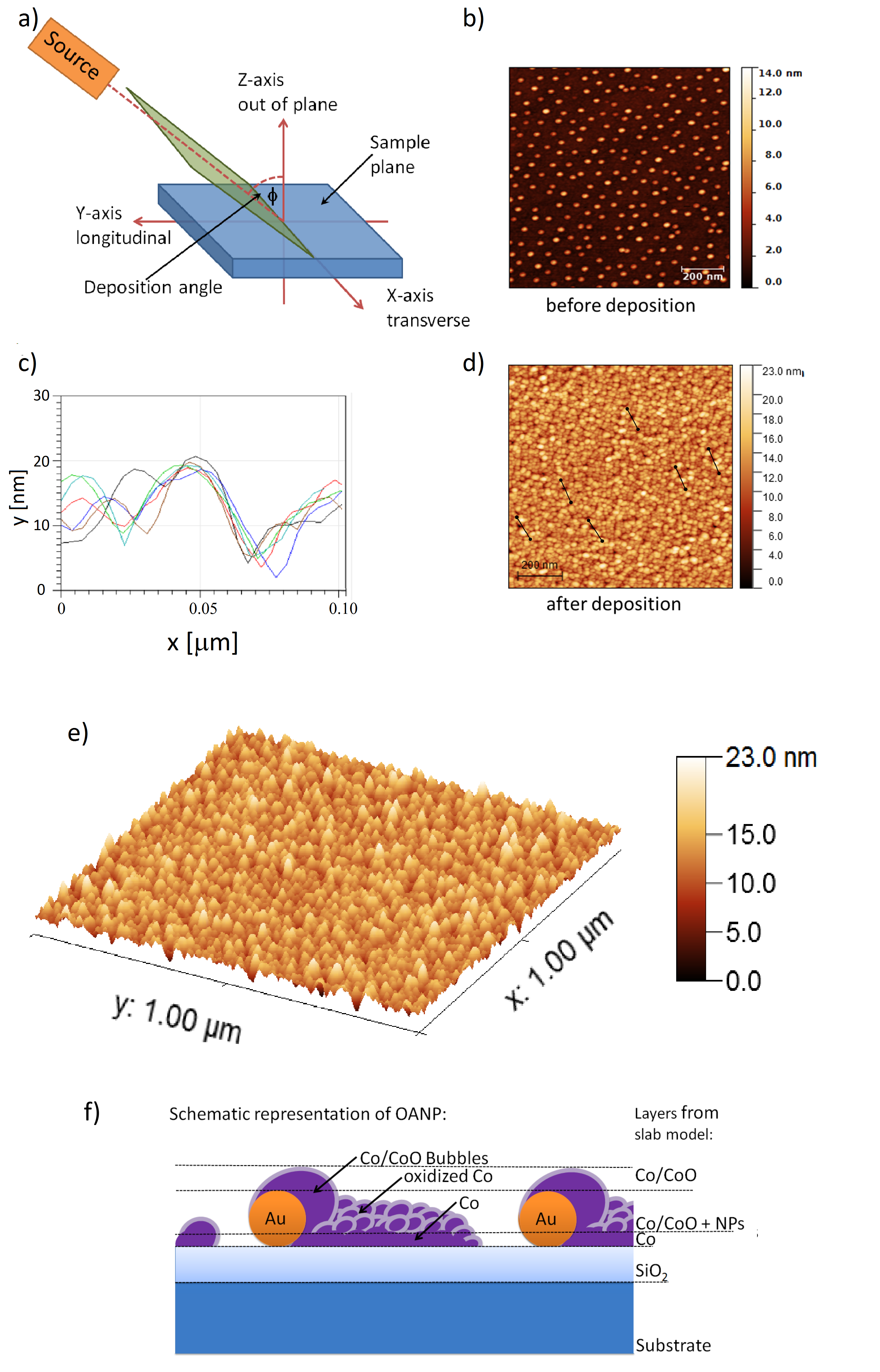}
	\caption{(a) Sketch of the sample orientation and all relevant directions; Atomic force microscopy images of OANP before (b) and after (d) the oblique angle deposition of nominally 20\,nm of cobalt; (c) depth profile of the indicated line segments in image (d) running across one ``bubble'' each; (e) 3~D representation of (d) emphasizing the structure of the film; (f) schematic representation of sample OANP, where the different subsystems as well as the layers used in the slab model are indicated (the relative thicknesses are not drawn to scale);}
	\label{fig:Fig1}
\end{figure}

The micellar \cite{Ethirajan2013} technique was used to decorate SiO$_2$ terminated Si(111) substrates with Au nanoparticles in a triangular pattern. The particle diameter ranges from 7-9~nm with an interparticle distance of 90-100~nm. 
The distance between the nanoparticles on the substrate was chosen such as to break up magnetic domain structures, which have been found to be larger or equal than 200~nm \cite{Girgis2003}. Co was deposited simultaneously on a nanopatterned sample (OANP) and a control sample (300~nm SiO$_2$ on Si(001)) without nanoparticles (OA) under an oblique angle of 81$^\circ$ using an electron gun. The electron gun is mounted under an angle of 26$^\circ$, with respect to the normal of a standard sample holder. To achieve a total deposition angle of 81$^\circ$ the samples were mounted on a 55$^\circ$ tilted wedge. Based on the geometry, we refer to the y-axis as longitudinal direction, while the x-axis, perpendicular to the direction of growth is referred to as transverse direction (Fig. \ref{fig:Fig1}a)).    

Taking into account the oblique incidence angle (see Fig. \ref{fig:Fig1}) according to $d = d_{nom} \frac{\cos(81^\circ)}{\cos(26^\circ)}$, the deposited thickness was $d = 3.5$~nm for a nominal thickness $d_{nom} = 20$~nm calibrated under the usual incidence angle of 26$^\circ$.
A CoO surface is formed by natural oxidization in air at room temperature. The surface structure of the samples was investigated using an AFM Multimode 8 System (Bruker, USA). The chemical layer structure of the samples was investigated using x-ray reflectometry (XRR) on a Bruker D8 Discover diffractometer with a Cu-source. Combined structural and magnetic depth profiles of the samples were obtained by polarized neutron reflectometry (PNR), performed on the D17 instrument at the Institut Laue-Langevin in Grenoble, France \cite{Cubitt2002, Saerbeck2018a}. The data is available via the data repository of the Institute Laue-Langevin \cite{dataILL}. The neutron reflectometry measurement was recorded without spin analysis in time-of-flight mode of operation using wavelengths from 4-20~\AA. A 7~T vertical Oxford cryomagnet was used to apply fields up to $\pm$~2~T and to reach the measurement temperature of 10~K. The magnetic field was applied along the sample plane, normal to the scattering plane. Prior to each measurement, a calibration measurement was performed to ensure the beam polarization was stable in each measurement condition. Both XRR and PNR data were fitted simultaneously using the GenX software package \cite{Bjo2007}. 
The fitting of reflectometry data is typically done using a model system of defined slabs with finite roughness, representing the scattering potential distribution as a function of depth. Within the nanoparticle system considered here, this approximation fails as the height modulations of the nanoparticles exceed the nominal film thickness of Co. The lateral coverage of Au nanoparticles is only about 1 \% assuming hexagonal arrangement of particles, which leads to a highly inhomogeneous lateral coverage of the sample. PNR is a lateral averaging technique and the experimental resolution applied in the experiment leads to a coherence area, the area over which reflected neutron waves interfere, of about 30 $\mu m$ along the neutron beam and a few Å in the transverse direction. The reflectivity of each coherence volume is averaged incoherently resulting in the measured total reflectivity. Due to lateral averaging exceeding the length scale of the nanoparticle separation, such lateral inhomogeneities are difficult to quantify from specular reflectometry alone. A functional model SLD profile based on, for example, Gaussian functions introduces many new parameters that may lead to an overdetermined fit. Therefore, we continue using the slab model but allow the roughness error functions to extend beyond the individual layer thickness to modulate the overall density with overlapping error functions. While this may lead to unphysical roughness parameters, the derived SLD profile can be taken as an approximate physical SLD profile as a function of depth. The bulk magnetic properties of the films were investigated using a LOT-Quantum Design MPMS3 SQUID VSM. 

\section{Results}

Atomic force microscopy (AFM) images of the patterned substrate before and after deposition are shown in Figures \ref{fig:Fig1} b) and \ref{fig:Fig1} d) respectively. Figure \ref{fig:Fig1} e) shows a 3D representation of the sample after deposition. After deposition, the overall sample surface shows a rough profile with height variations of about 10 nm. Additionally, larger structures with height variations of 20 nm can be distinguished. The latter arises from Co/CoO ``bubbles'' forming around the Au NPs. Consequently, we separate the structure into two subsystems, subsystem 1 containing the exchange biased, granular Co/CoO film between the NPs and subsystem 2 containing the Co/CoO ``bubbles'' around the NPs. 

Figure \ref{fig:Fig1} f) shows a schematic view of sample OANP highlighting the different subsystems as well as the layers defined in the slab model which was used to fit the PNR data.
Figure \ref{fig:Fig1} c) shows the depth profile across several ``bubbles'' indicated in Figure \ref{fig:Fig1} d). 
From these profiles ``bubbles'' and the shadows behind them can be seen more clearly.

\begin{figure}
	\centering
		\includegraphics[width=1.0\textwidth]{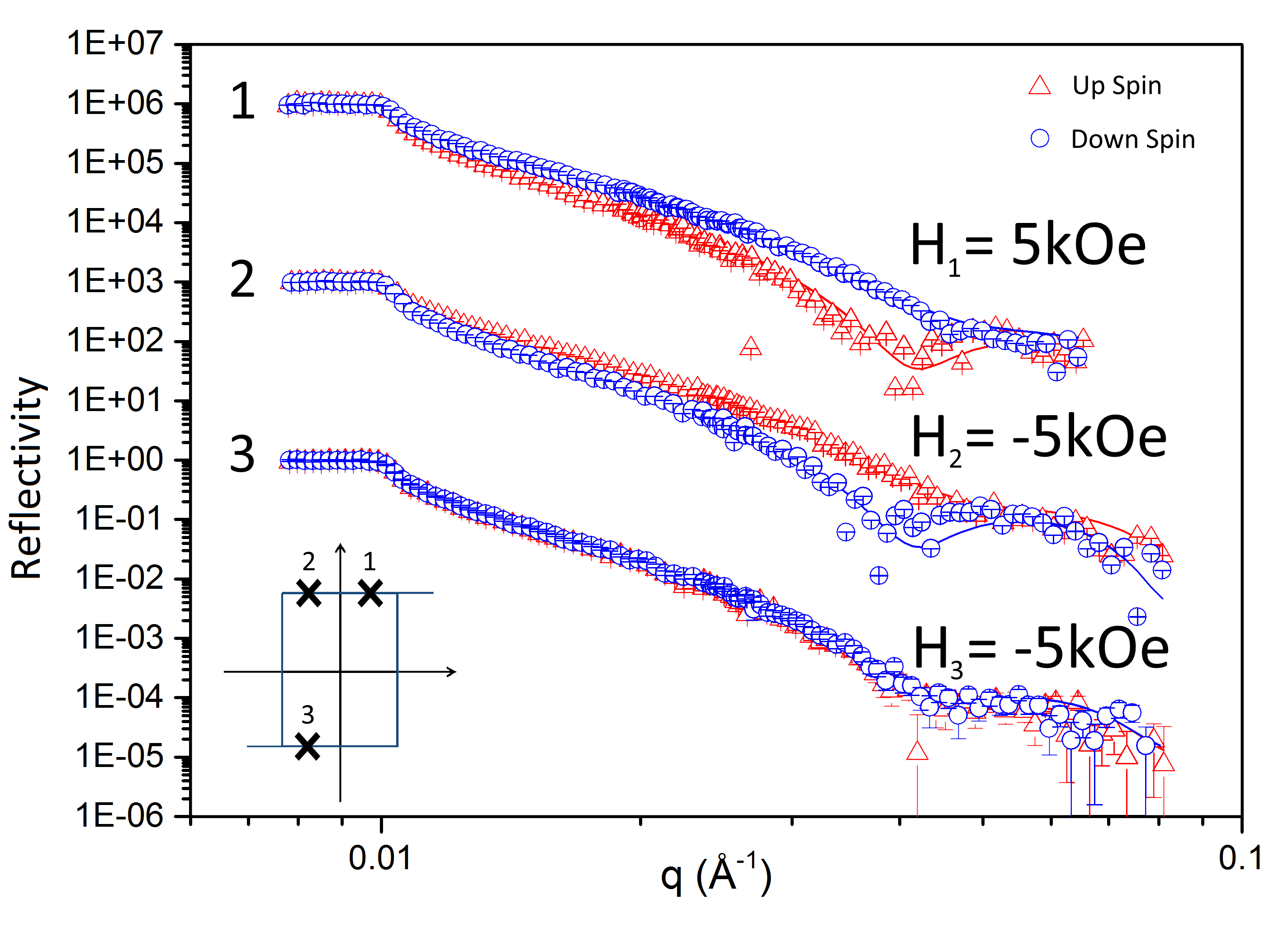}
	\caption{PNR data of sample OANP. Red open triangles show the reflectivity R+ with neutron spin parallel to the external field. Blue open circles show the reflectivity R- for which neutron spin state and external field are antiparallel. The full lines in the corresponding colour represent the fit. Curves 1 and 2 have been offset by 10$^{3}$ counts for clarity. The inset shows the fields along the hysteresis loop where each data set was recorded.}
	\label{fig:Neutron_data3}
\end{figure}

\begin{figure}
	\centering
		\includegraphics[width=0.7\textwidth]{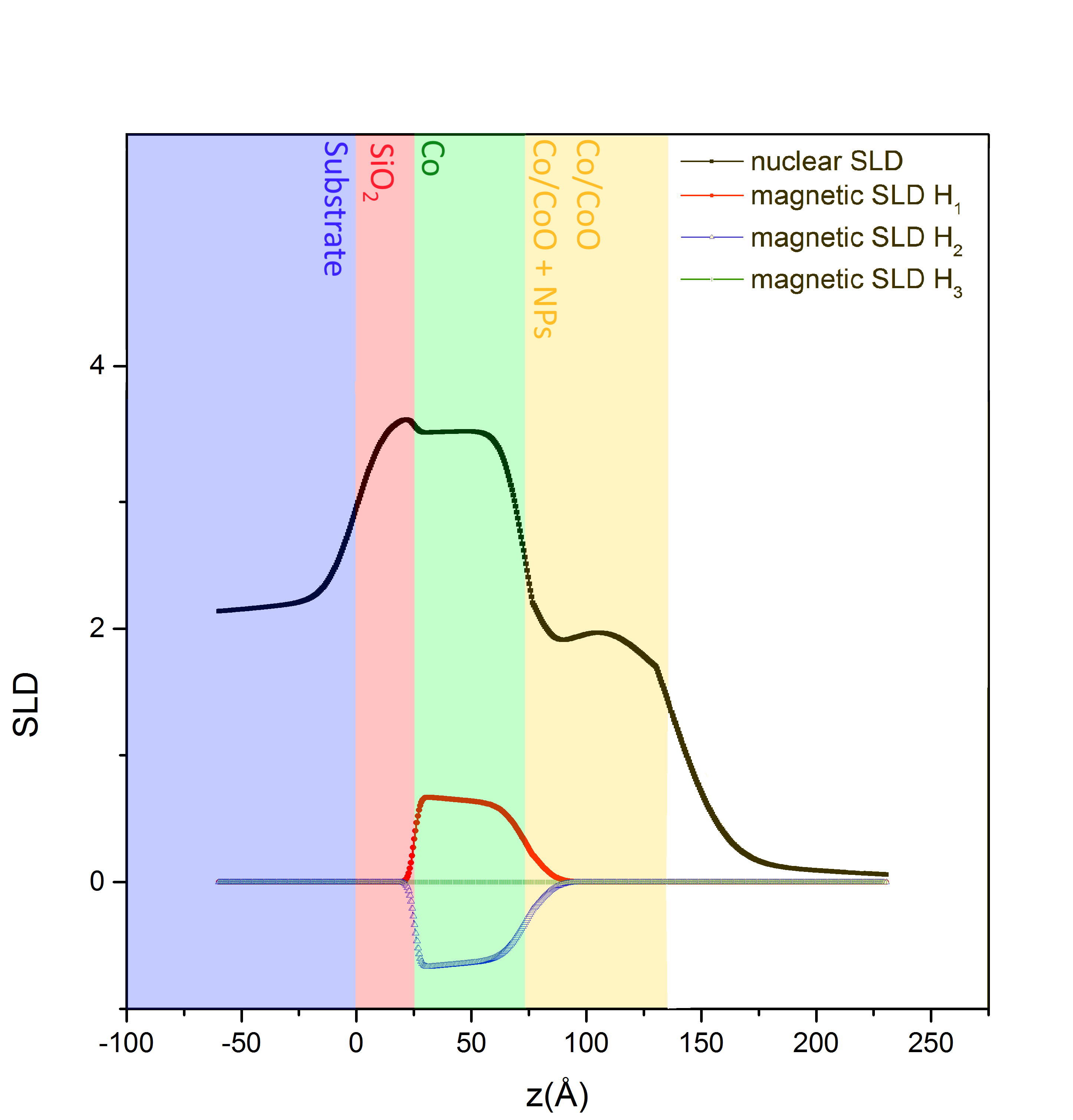}
	\caption{Scattering length density (SLD) extracted from the fit of the PNR data shown in Figure \ref{fig:Neutron_data3}. The nuclear SLD is the same for all three data sets whereas the magnetic SLD changes as a function of field. The inset shows a sketch of the sample structure. The slabs corresponding to the different layers in the model SLD are indicated using different background colours.}
	\label{fig:SLD_reduced2}
\end{figure}

The depth resolved structure and distribution of the long-range ordered magnetic moment was elucidated with specular PNR at 10~K after field cooling in 5~kOe. Figure \ref{fig:Neutron_data3} shows the reflectivity obtained from OANP at different external magnetic fields along the hysteresis curve as indicated in the inset. The data were fitted using the magnetic reflection model in the GenX  software \cite{Bjo2007}, which is based on the recursion matrix algorithm introduced by Stepanov and Sinha \cite{Stepanov2000}. We applied a standard slab model approach with a sliced scattering length density of 0.5~\AA~intervals. PNR is a laterally averaging technique, which does not allow direct access to the individual nanoparticles on these length scales. Instead, the composition and magnetization are laterally averaged at each depth. A model of the sample structure resembling the SLD profile is shown in the inset of Figure \ref{fig:SLD_reduced2}. We did not observe significant off-specular scattering above background from the arrangement of nanoparticles within the aquisition time used for the specular measurements. Due to the small length scales involved, the low sample volume and irregularities in the nanoparticle ordering, possible off-specular signals will be weak and spread significantly over the 2D area detector. 

The fitting provides an approximate depth profile of nuclear and magnetic densities (Figure 3). We determine a magnetic layer of 49~\AA~closest to the substrate and two largely smeared non-magnetic layers of 37~\AA~and 23~\AA~representing the Co/CoO granular film on top of the Co-layer and the ``bubbles'', respectively (compare Figure 1f). The thicknesses of the layers should not be taken literally as the layers are merged by the extended roughness required by the slab model. Due to the large roughness, the two top layers are also not indicated by different colors in Figure 3. Nevertheless, the nuclear and magnetic depth information extracted from the fits to the PNR data accurately correlates the location of the magnetic layer with respect to the structural profile. The SLD profiles further suggest that subsystem 1 needs to be further sub-divided into a magnetic Co layer and a granular Co/CoO structure, the latter being merged with the ``bubbles'' as a function of depth.
The best match to the data is obtained when the magnetization is confined to a region close to the substrate, which would agree with a Co layer of larger lateral dimension being formed between the particles. We do not detect any measurable magnetization located in the surface structures, or ``bubbles''. Due to the coherent lateral averaging of PNR on the length scales of the NP separation, one can expect that magnetic features of nanoparticles are averaged and the reflectivity is dominated by the more continuous thin film parts of the sample. After field cooling, the measurement at +5~kOe shows a magnetic splitting corresponding to 0.35~$\mu_B$/atom. The magnetic splitting was calculated from the magnetic SLD using the bulk density for cobalt, 8.86~g/cm$^{3}$ corresponding to 0.09~atoms per \AA$^{3}$. This value corresponds to approximately half of the value from SQUID magnetometry: 0.61~$\mu_B$/atom. The latter was calculated by dividing the magnetic moment at 5~kOe and 10~K, after background subtraction, by the number of atoms in the sample (Bulk: $\approx$ 1.7~$\mu_B$/atom \cite{Coey2010}), and a measured sample area of 24.15 mm$^2$. We observed a strong variation of the high-field slope of the magnetization curves as a function of temperature, which makes an estimate of the actual background difficult. The difference in magnetization between SQUID magnetometry and PNR is most likely due to the fact that PNR is mostly sensitive to long-range ordered moments, while SQUID magnetometry is sensitive to every magnetic moment in the volume, including paramagnetic and uncompensated antiferromagnetic spins. For the estimate of the magnetization in units of $\mu_B$/atom measured by SQUID, the density of the Co layer plays a crucial role. In a disordered, granular system with different degree of oxidation, a large discrepancy between assumed and actual atomic density can occur, which can make the obtained moment unreliable. Indeed, PNR does reveal a scattering length density which is not constant over the entire thickness, but the measurement does not allow to separate the atomic density from atomic species but only measures the product of average atomic scattering length and unit volume. Therefore, we provide the estimated magnetization based on the bulk Co density.\\ 
Upon reduction of the field to -5~kOe, the splitting in R$^+$ and R$^-$ reverses in the PNR measurement, which indicates an antiparallel alignment of sample magnetization and external magnetic field. The absolute magnetization did not decrease within the 1~\% accuracy of the measurement, indicating that -5~kOe lies well within the coercivity of this part of the sample. The measurement at -5~kOe was repeated after negative saturation in -20~kOe. The splitting between up and down spins collapses for this measurement point, which implies zero averaged magnetization in the measurement plane and indicates a non-collinear domain state.

\begin{figure}
	\centering
		\includegraphics[width=1\textwidth]{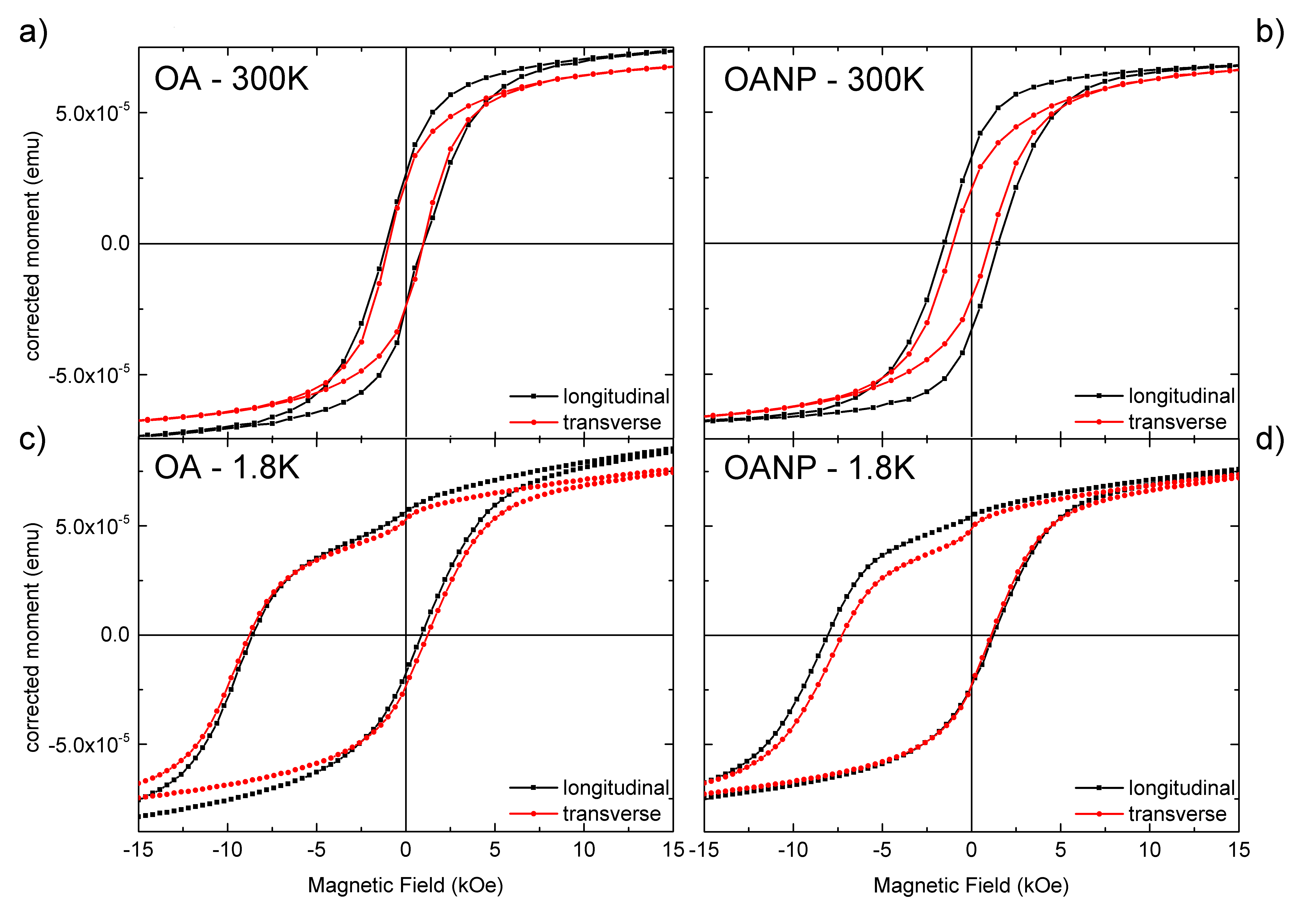}
	\caption{Magnetic hysteresis loops of samples OA and OANP recorded at 300~K as well as 1.8~K, for the longitudinal and transverse directions}
	\label{fig:hysteresis}
\end{figure}

\begin{table}
	\centering
	\caption{Coercive Field and exchange bias, for the sample with (OANP) and without NPs (OA), longitudinal and transverse to the growth direction at 1.8~K; Coercive field and Exchange bias have been calculated using: $H_C = \frac{1}{2}(H_{C_2} - H_{C_1})$ and $H_{EB} = \frac{1}{2}(H_{C_2} + H_{C_1})$} 
		\begin{tabular}{|c | c | c |c | c |}
		\cline{1-5}
			\pbox{4cm}{ \multirow{2}{*}{sample and direction }}	& \pbox{2cm}{H$_{C_1}$} & \pbox{2cm}{H$_{C_2}$} &	\pbox{2cm}{H$_C$}	& \pbox{2.5cm}{H$_{EB}$} \\ 
		& [kOe] & [kOe] & [kOe] & [kOe] \\
		\cline{1-5}		
		OANP long. & -8.1	&	1.2 &	4.6 & -3.4	 \\
		\cline{1-5}
		OANP transv. & -7.3	&	1.1 &	4.2  & -3.1  \\
		\cline{1-5}
		OANP $\Delta$(long.-transv.) & $-$	&	$-$ &	0.5  & 0.3  \\
		\cline{1-5}
		OA longitudinal & -8.6 & 0.9	& 4.7 & -3.9  \\
		\cline{1-5}
		OA transverse & -8.8 &	1.2 & 5.0 &	-3.8	\\
		\cline{1-5}
		OA $\Delta$(long.-transv.) & $-$	&	$-$ &	0.3  & 0.1  \\
		\cline{1-5}
		\end{tabular}
	\label{tab:HC}
	\end{table}

Figure \ref{fig:hysteresis} compares the untrained magnetic hysteresis (after background subtraction) of the samples measured along the longitudinal and transverse direction with respect to the growth direction at 300~K and 1.8~K after field cooling in 5~kOe. 
For all measurements the field and measurement directions are equal. 
The figure shows the hysteresis curves in a range from -15~kOe to +15~kOe to illustrate the shape of the loops around the coercive fields more clearly. The samples reach saturation at approximately 25~kOe, where the difference in magnetic moment disappears. For sample OA there exists a small discrepancy in saturation magnetization between the two measurement directions which most likely stems from a difference in stray fields in the different orientations, which can lead to a difference in magnetic moment in SQUID magnetometry \cite{QD1014201, QD1500015}.
Without nanoparticles, at 300~K, no clear anisotropy can be identified with both directions showing essentially the same relative remanence of 32~\% of the saturation value and coercive field values of about ($200\pm10)$~Oe (Figure \ref{fig:hysteresis}a). With nanoparticles (Figure \ref{fig:hysteresis}b), the longitudinal direction shows a more rectangular hysteresis, resulting in a higher remanent magnetization M$_{rem}$ of 44~\% of the saturation value, as compared to 29~\% for the transverse direction. The coercive field increases in the longitudinal direction to H$_C$ = 300~Oe, while H$_C$ = 208~Oe is observed with the field applied transverse to the growth direction. This indicates a growth induced in-plane anisotropy only for the OANP case. Because these measurements are taken above the Néel temperature of CoO, no exchange bias is expected.

\begin{figure}[h]
	\centering
		\includegraphics[width=1\textwidth]{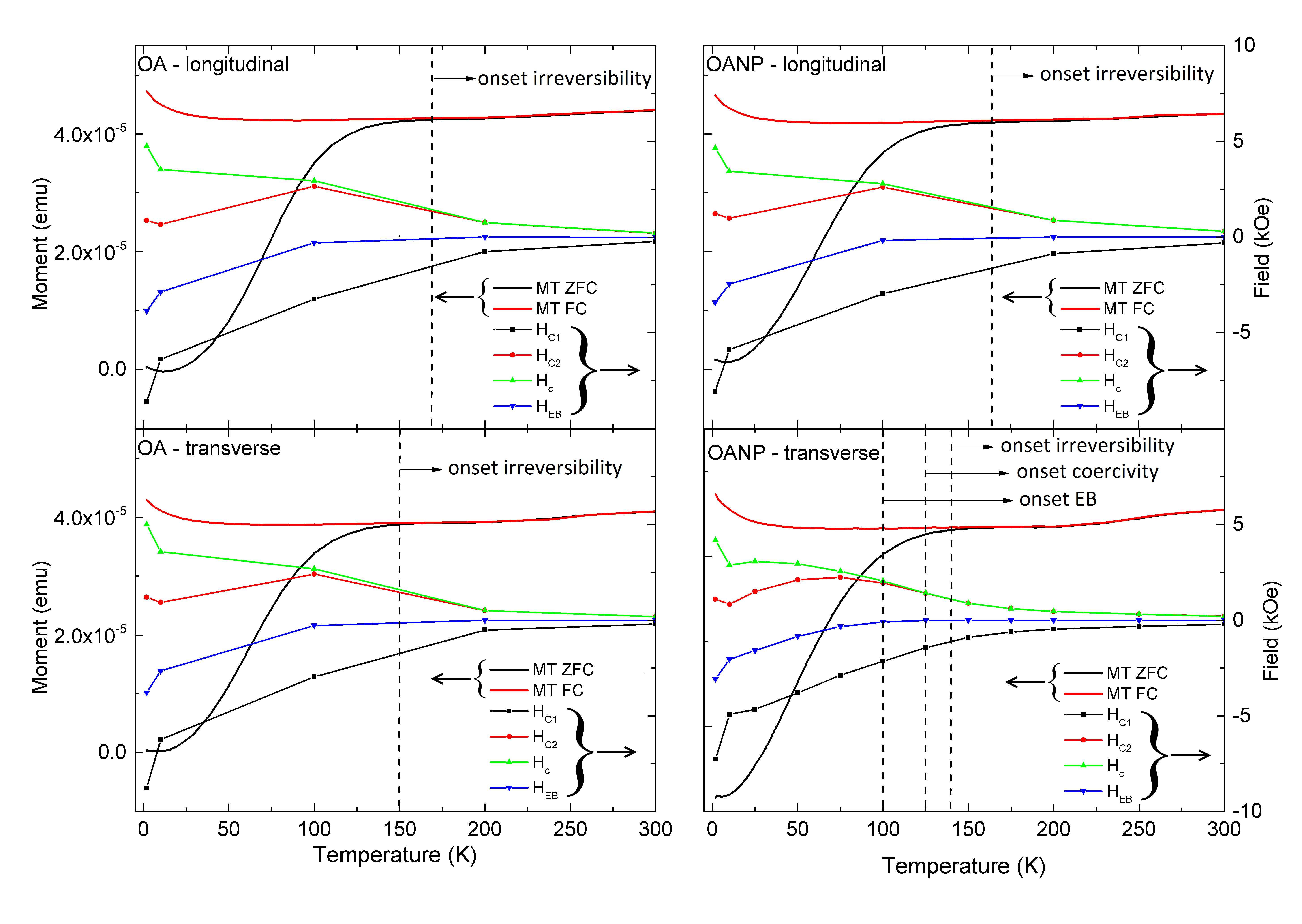}
	\caption{Temperature dependence of the magnetization, coercive fields, and exchange bias for the longitudinal as well as the transverse direction for samples OA and OANP. The left hand side of the Y-axis corresponds to the magnetic moment in emu whereas the right hand side shows fields in kOe.}
	\label{fig:TempBehaviour}
\end{figure}

Upon field cooling to 1.8~K in either longitudinal or transverse direction, both samples exhibit strong exchange bias, $H_{EB}$ = $\frac{1}{2}(H_{C_2} + H_{C_1})$ and increased coercive fields, $H_{C}$ = $\frac{1}{2}(H_{C_2} - H_{C_1})$ (Figures \ref{fig:hysteresis}c) and \ref{fig:hysteresis}d)). Additionally, a small kink at low fields can be observed.
Such a kink has been observed before in different Co/CoO nanostructures \cite{Temst2006, Popova2005, Dobrynin2006}. It has been proposed that thin polycrystalline Co/CoO nanostructures consist of a Co/CoO exchange bias system where the ferromagnetic spins are pinned to the antiferromagnetic domains, as well as a purely ferromagnetic cobalt layer that is free to switch magnetization independently, therefore leading to a small kink at low reversal fields. Such a two component system agrees well with the further sub-division of subsytem 1 as derived from the PNR fits.  
The kink is more pronounced for the transverse than the longitudinal direction. 

The temperature dependence of the magnetization, coercive fields and exchange bias is summarized in Figure \ref{fig:TempBehaviour}. The differences in temperature dependence of magnetization, H$_C$, and H$_{EB}$ between the longitudinal and transverse orientation as well as OA and OANP are marginal. The onset of irreversibility, i.e. the point where field cooled and zero-field cooled magnetizations split, is 20~K higher along the longitudinal direction for both samples. 
The evolution of the coercive fields (H$_{C_1}$, H$_{C_2}$, H$_{C}$) and the exchange bias (H$_{EB}$) of both orientations is very similar as well. It can however be observed from Table \ref{tab:HC} and Figure \ref{fig:hysteresis} that for sample OANP the values for H$_{C_1}$, H$_{C_2}$, H$_{C}$ and H$_{EB}$ are consistently higher for the longitudinal direction, while for sample OA the differences in H$_{C}$ and H$_{EB}$ are smaller. 
This increased coercivity can indicate additional domain wall pinning or anisotropy for the longitudinal direction. The reduced coercivity in the transverse direction on the other hand could be originating from a hard axis contribution to the hysteresis loop which shears the loop.

\section{Discussion}

A remarkable difference is observed between moments obtained from SQUID hysteresis loops and PNR measurements with different field history.

\begin{figure}[h]
	\centering
		\includegraphics[width=0.80\textwidth]{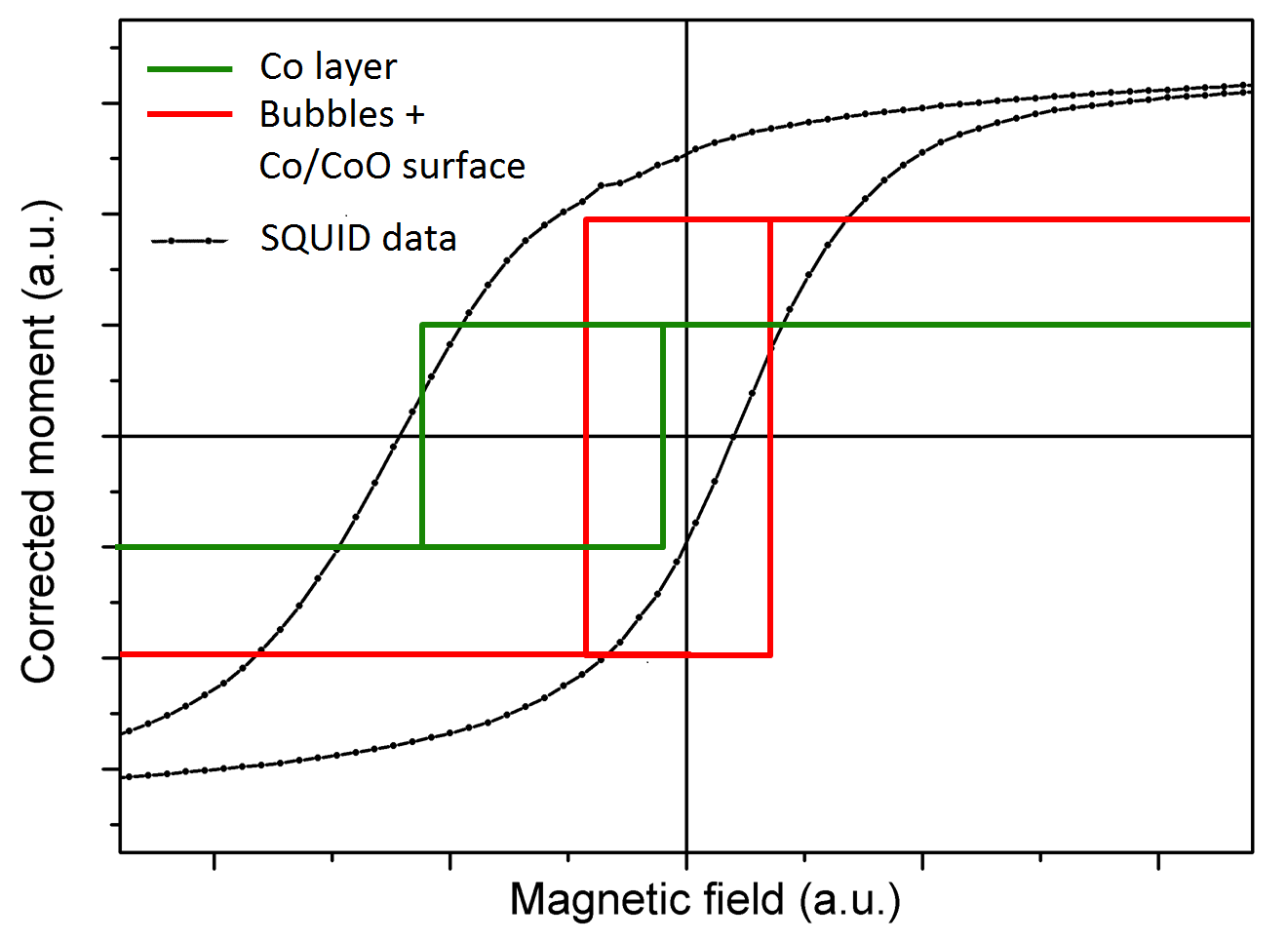}
	\caption{A sketch of the hysteresis loops of the different subsystems plotted over the hysteresis loop measured with SQUID magnetometry at 10~K. }
	\label{fig:PNR_SQUID}
\end{figure}

A model of two surface subsystems is supported by the AFM images after deposition in Figure \ref{fig:Fig1} d) and e):
an exchange biased, granular Co/CoO film covering the area between the Au NPs makes up subsystem 1 while the Co/CoO ``bubbles'' that have formed around the Au NPs make up subsystem 2.
The PNR data suggest a further sub-division of subsystem 1 into a Co layer close to the substrate with long-range ordered moments and a granular Co/CoO layer closer to the surface. The antiferromagnetic CoO of this surface layer provides the pinning for the measured exchange bias. 
In positive saturation (point 1, Figure 2 inset and Figure 6), the fit provides a structural profile and the depth distribution of the magnetic moment, which is concentrated in a layer close to the substrate. Upon reversal of the field to -5 kOe, the moment in this layer does not change sign or decrease significantly, which shows that the applied negative field is above the field-decreasing coercive field of this layer (point 2 Figure 2 inset and Figure 6). In contrast, when the same negative field is approached from negative saturation (point 3 Figure 2 inset and Figure 6) zero averaged magnetization is detected. This suggests that -5 kOe is close to the field-increasing coercive field value of this layer. This hysteresis behaviour of the Co layer close to the substrate is illustrated as “Co layer” in Figure 6. 
The larger magnetic moment observed in SQUID magnetometry (at point H$_1$ of the hysteresis loop) indicates that a significant amount of magnetic moments is not detected by PNR. The increased disorder and inhomogeneous oxidation in the granular surface of subsystem 1 and the Co/CoO ``bubbles'' of subsystem 2 leads to a highly diluted magnetic system. Since the features of the magnetic structure are averaged over the coherence length of 30 $\mu$m, the resulting moment from these two components may be too diluted to be detected in PNR.  Further lateral variations on length scales larger than the coherence volume may additionally mask any magnetic signal. Therefore, the magnetic signal in PNR is dominated by the more continuous Co-patches close to the substrate, which corresponds to approximately 50 \% of the total moment estimated from the SQUID data. 
At point H$_2$ of the hysteresis curve, SQUID measurements show a drastically reduced moment, indicating that the “bubbles” (subsystem 2) and the granular Co/CoO surface of subsystem 1 have a much reduced coercive field. Furthermore, a smaller decrease in moment is detected by SQUID in the field increasing -20~kOe $\rightarrow$ -5~kOe branch, which indicates the field is lower than the coercive field. This indicates a system with smaller exchange bias and narrow hysteresis, which is illustrated as “bubbles + Co/CoO surface” in Figure 6.
Indeed, one can expect a smaller exchange bias and narrower hysteresis in the granular Co/CoO layer and the ``bubbles'', due to an increased anti-ferromagnet to ferromagnet ratio and, due to the large disorder and a large amount of uncompensated moments acting more similar to paramagnetic spins. The hysteresis of the granular Co/CoO and the ``bubbles'' is schematically illustrated by subsystem 2 in Figure \ref{fig:PNR_SQUID}.

The different degrees of oxidation and therefore different values for the exchange bias, will lead to a very spread out and rounded hysteresis. SQUID measures the volume magnetization and therefore cannot distinguish between the three magnetic systems, the Co-layer close to the substrate, the granular Co/CoO structure close to the surface and the ``bubbles'' formed around the Au NPs, which are combined in the hysteresis in Figure \ref{fig:PNR_SQUID}.

While for sample OA no magnetic easy axis can be determined, for sample OANP the magnetic easy axis lies along the longitudinal direction.
This difference is likely the result of the nanoparticle decoration. The nanoparticles seem to enhance the anisotropy in the OAD film, by increasing the shadowing effects during early stages of OAD growth.

\section{Conclusion}

In conclusion, we have shown the presence of exchange bias in an ultra thin Co/CoO film grown under OAD on a nanostructured substrate. We found that the nanoparticles add another variable to the interplay between layer thickness and deposition angle that determines the direction of the magnetic easy axis in OAD magnetic nanostructures. For cobalt deposited under an angle of 81$^\circ$ the nanoparticles lead to a magnetic easy axis along the longitudinal direction, whereas for the sample without nanoparticles no preferred magnetization direction was observed.
AFM and PNR suggest a separation of the structure into three different substructures, a Co-layer close to the substrate, a granular Co/CoO surface and ''bubbles'' forming around the nanoparticles. The combination of SQUID and PNR measurements at different applied fields allows to disentangle the magnetic response of each system to a certain degree, revealing strong exchange bias in the Co-layer and a narrow hysteresis with little to no exchange bias in the remaining two systems.    

\section{Acknowledgments}

We acknowledge the financial support from the Research Foundation Flanders (FWO), the Concerted Research Action GOA/14/007 and C14/18/074 of KU Leuven, the Hercules Foundation and the MagNet Project. The authors want to thank Ambroise Peugeot for his help to prepare and his participation in the PNR measurements.

\bibliography{allpapers}

\end{document}